\documentclass[mathleft
]{an}
\usepackage{graphicx}
\usepackage[sorting=none]{}
\usepackage{dblfloatfix} 
\usepackage{fixltx2e}
\usepackage{times}
\usepackage{enumerate}
\usepackage{url} 
\makeatletter
\def\url@leostyle{%
  \@ifundefined{selectfont}{\def\UrlFont{\sf}}{\def\UrlFont{\footnotesize\ttfamily}}}
\makeatother
\usepackage{hyperref}
\usepackage[hyphenbreaks]{breakurl}
\usepackage{natbib}
\bibpunct{(}{)}{;}{a}{}{,}
\begin{document}

% The following seven commands are intended for editorial usage and should be ignored by
% the author(s).
\Pagespan{001}{}% Document's page range. 
% If second parameter is left empty, the last page is computed automatically.
\Yearpublication{2012}%
\Yearsubmission{2012}%
\Month{xx}%   
\Volume{xxx}%  
\Issue{xx}% 
% \DOI{This.is/not.aDOI}% 

\title{Principles of High-Dimensional Data Visualization in Astronomy}

\author{Alyssa A. Goodman\inst{1}\fnmsep\thanks{Corresponding author:
  \email{agoodman@cfa.harvard.edu}\newline}
%Example 
%for footnote, note the usage of the \texttt{fnmsep}
%command as separator between institute number and footnote mark} 
}
\titlerunning{Astronomical Visualization}
\authorrunning{A. Goodman}
\institute{
Harvard-Smithsonian Center for Astrophysics, Cambridge, MA, USA}

\received{20 Apr 2012}
\accepted{20 Apr 2012}
\publonline{later}

\keywords{techniques: image processing; methods: data analysis; techniques: radial velocities; cosmology: large-scale structure; ISM: clouds}

\abstract{%
 Astronomical researchers often think of analysis and visualization as separate tasks.  In the case of high-dimensional data sets, though, interactive {\it exploratory data visualization} can give far more insight than an approach where data processing and statistical analysis are followed, rather than accompanied, by visualization. This paper attempts to charts a course toward ``linked view" systems, where multiple views of high-dimensional data sets update live as a researcher selects, highlights, or otherwise manipulates, one of several open views.  For example, imagine a researcher looking at a 3D volume visualization of simulated or observed data, and simultaneously viewing statistical displays of the data set's properties (such as an $x$-$y$ plot of temperature vs. velocity, or a histogram of vorticities).  Then, imagine that when the researcher selects an interesting group of points in any one of these displays, that the same points become a highlighted subset in all other open displays.   Selections can be graphical or algorithmic, and they can be combined, and saved.  For tabular (ASCII) data, this kind of analysis has long been possible,  even though it has been under-used in Astronomy.  The bigger issue for Astronomy and several other ``high-dimensional" fields  is the need systems that allow full integration of images and data cubes within a linked-view environment.  The paper concludes its history and analysis of the present situation with suggestions that look toward cooperatively-developed open-source modular software as a way to create an evolving, flexible, high-dimensional, linked-view visualization environment useful in astrophysical research.
 }

\maketitle

\section{Introduction}
Historically, Astronomy has been a visual science.  Thousands of years ago observations were carried out with the naked eye; hundreds of years ago telescopes  augmented the eye; and during the last century sensitive film and CCD recording devices enhanced what the eye could see.  More recently, observing techniques spanning the full electromagnetic spectrum have been developed, as have techniques for statistical comparison with  analytic and numerical theoretical predictions.  Oddly though, as Astronomy's  wavelength coverage increased, the value of the ``visual" to astronomers seems to have declined--not as a wavelength, but as a tool.   Too often, wavelength-specific studies of tiny patches of sky, or statistical analyses of tremendous catalogs of information, are carried out with very little attention paid to context.  Viewing what {\it surrounds} a tiny narrow-field image, or studying a catalog's content in context on a wide-field sky often gives unexpected and valuable information.  Understanding the context of catalog data in  high-dimensional spaces where information can be compared across wavelengths and across models, can be similarly illuminating.  Evolution has made humans amazingly good at pattern recognition, and this paper is about how analysis techniques that marry humans' extraordinary visualization capabilities to statistical principles are, and should continue to be, on the rise within modern astronomy.\footnote{\citet{HassanAmr2011} recently published  a uniquely comprehensive review of the recent history of visualization in Astronomy, and the interested reader is referred to that work for details and links to software not provided here.}
\begin{figure*}
% \vspace*{-2.0 cm}
\begin{center}
 \includegraphics[width=6.5in]{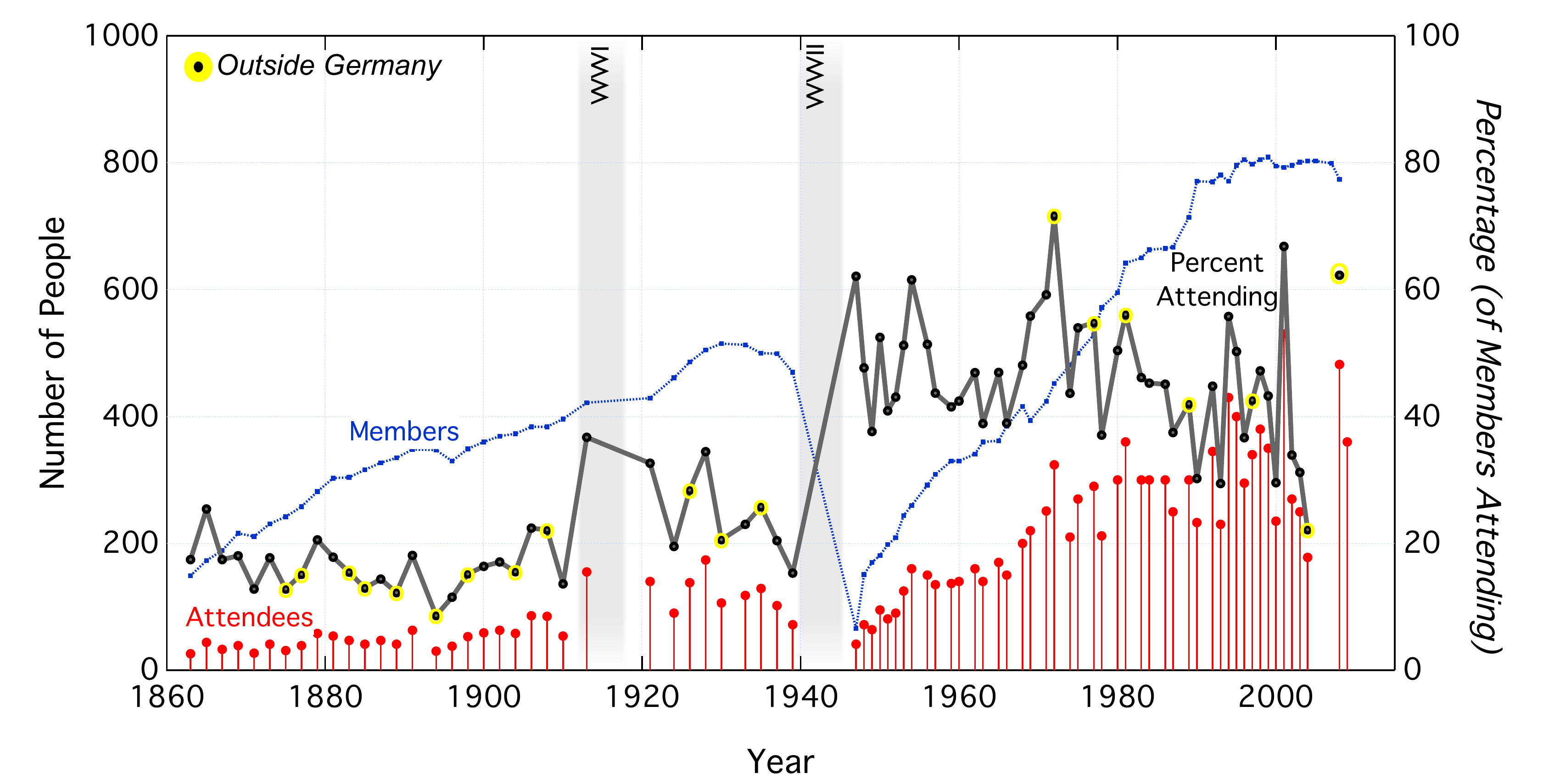} 
 \end{center}
\vspace*{0.2 cm}
\caption{History of the AG meeting.}
\label{history}
\end{figure*}
   
\section{Data-Dimensions-Display}

There are three simple words to keep in mind when one sets out to explore and/or explain high-dimensional information with visualization: {\bf data}, {\bf dimensions}, and {\bf display}.  Any {\bf data} set containing the equivalent of more than two columns worth of information can be thought of as ``high-dimensional."  In some cases, the {\bf dimensions} may be spatial or temporal, but in other cases the dimensions might be just columns in a data table, so a ``high-dimensional" space can be highly abstract.\footnote{\citet{Wong1997} provide an excellent review of multi-dimensional multi-variate visualization that includes a good discussion of the meaning of the word ``dimensions" within various disciplines.}

Consider Figure \ref{history}, which shows a simple cartesian graph documenting attendance at  {\it Astronomische Gesellschaft} (AG) meetings over time.  The {\bf data} used to create this graph are from the AG website,\footnote{\url{www.astronomische-gesellschaft.org/en/tagungen}} which contains a table with 8 columns, listing: Year, RGA\footnote{An index number on ``regular" meetings of the General Assembly.}, City, Date, Number of Members, Number of Attendees, Number of Lectures, and  Number of Posters.   Thus, this {\bf data} set has at least 8 {\bf dimensions} (and more if locations' GPS coordinates were to be used in lieu of placenames).  

In order to best convey meaning to a particular audience, one needs to consider the {\bf display} mode that can and will be used.  For example, when I presented Figure \ref{history} at the 2011 AG meeting, to a group of astronomers, I used slideware on a data projector.   Even though I could have created and shown a kooky unconventional {\bf display} (e.g. a time-lapse movie showing a world map where spinning graphs representing ratios of attendees/talks/posters at AG meetings float above relevant cities), I knew that my audience would not expect or understand such a {\bf display}.  So, I  chose instead a standard $x$-$y$ time-series-style graph, where {\bf dimensions} ({\it number of members, number of attendees}) are plotted as number vs. time  and where a calculated diagnostic, percentage of members attending is shown using an additional  (right-hand) $y$ axis, but a shared $x$-axis (time). Partial information from one additional {\bf dimension} is also shown, since yellow-highlighted points indicate {\it locations} outside of Germany. Thus, {\bf 4+ dimensions} (three tabulated, one calculated, one partial) are shown in a {\bf 2-dimensional display}.  Context from beyond the online table is added to this display in the form of labeled grey bands showing the duration of the two world wars, which explain gaps in the series of meetings.  Subtle stylistic choices\footnote{The works of Edward Tufte \citep[e.g.][]{Tufte2001} are an excellent general resource concerning how to optimize visual displays of quantitative information.} about {\bf display} are also made, so that, for example, attendance numbers are shown as a series of vertical lines connecting dots to the zero-line, looking a bit like a histogram made of ``headed" symbols.  The graph is labeled within its borders, so as to avoid the need for an extensive caption.  

\begin{figure}[!h]
% \vspace*{-2.0 cm}
%\begin{center}
 \includegraphics[width=3.25in]{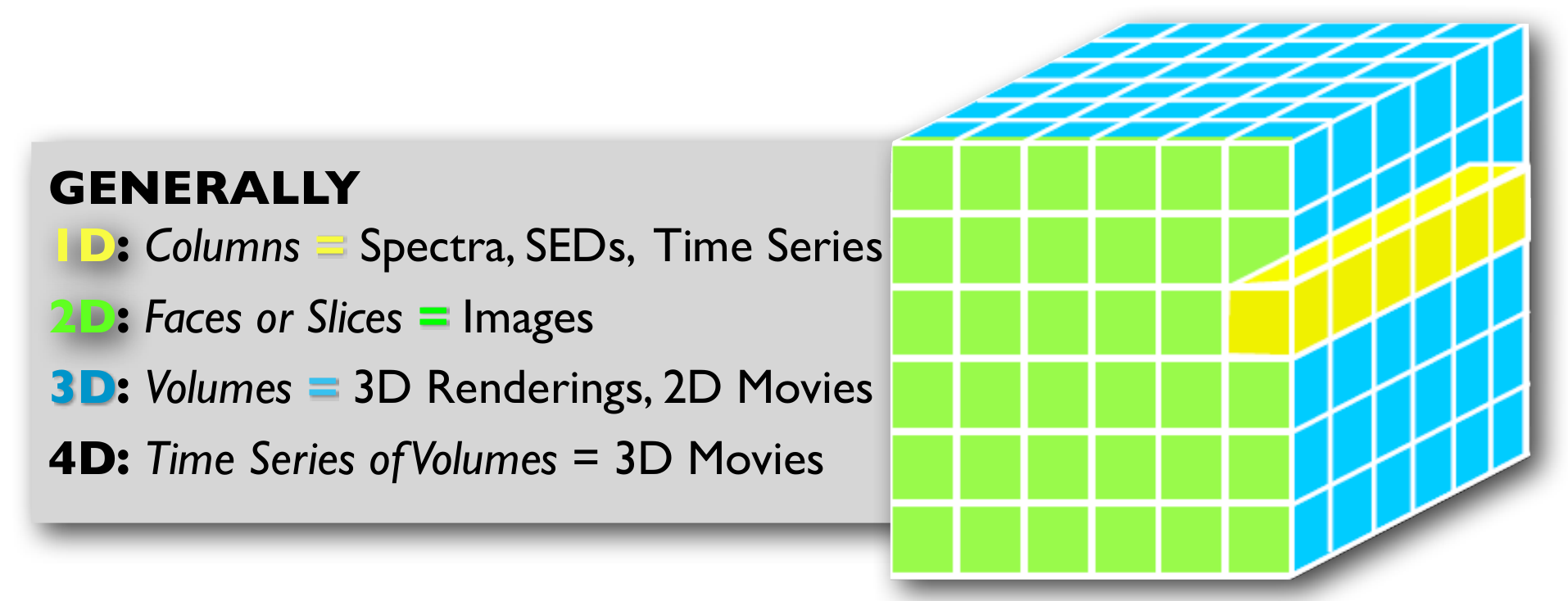} 
\vspace*{0.2 cm}
\caption{Data ``Cubes" in Astronomy.}
\label{datacubes}
\end{figure}

The AG meetings example in Figure \ref{history} offers a very specific, time-series-based, example of  Data-Dimensions-Display principles,  but deeper value is to be had when D-D-D is considered as a  more general construct.   Figure \ref{datacubes} shows a cube representing an abstract three-dimensional space.  In Astronomy, and most other sciences,   {\bf data} are often acquired as a function of many {\bf dimensions} (e.g. intensity as a function of space, time, wavelength, etc.).\footnote{In commercial data analytics and statistical analysis systems,  high-dimensional ``hyper-cubes" are commonly analyzed and visualized using ``OLAP" (online analytical processing) technologies.}  {\it But}, subsets of those data are usually only displayed and analyzed along one or two dimensions at a time (e.g. as a spectrum showing intensity as a function of wavelength).

Consider the color-coded examples listed in the grey box associated with Figure \ref{datacubes}.    In Astronomy, intensity as a function of {\it one} (non-spatial) dimension is most frequently thought of and displayed in an $x$-$y$ graph as a spectrum, an SED, or a time-series.  Intensity as a function of {\it two} (spatial) dimensions often is appropriately thought of and displayed as an image or contour map.  In many cases, such as in maps of spectral-line emission or layers of data at multiple wavelengths,  images or contour maps can be contextualized as ``slices" through a higher-dimensional ({\it 3D}) space that forms what is typically called a ``data cube."  

The set of {\bf display} modes  for seeing all the data in a cube is growing and presently features static 3D renderings, stereoscopic display, and interactive representations \citep{HassanAmr2011}.
In cases where it is possible to generate a series of data cubes as a function of some {\it fourth} dimension (usually time),  3D animations and/or sets of small multiples (repeated versions of 3D views seen side-by-side) are often used for display.\footnote{There are many excellent software packages capable of achieving beautiful visualizations of high-dimensional real and simulated data, nearly all of which are explained and listed in the recent review by \citet{HassanAmr2011}.  Here, I have chosen to focus instead on ideas about how to {\it link} the information in visualizations amongst otherwise-hidden dimensions and aspects of a data set.}

Software for analyzing observations and simulations is constantly growing more capable due to increased computational performance.  But, even the most modern astronomical software packages still do not make enough use of explicit connections between the dimensions inherent in a data set. Instead,  various kinds of displays ($x$-$y$ graphs, images, volume renderings) are created separately, using tools that do not link common dimensions across all active plots.  Below, I explain a ``linked view" approach that is likely to become the {\it essential} path to insight as astronomical data sets continue to expand in complexity and size.

\section{Linked Views}

\begin{figure}[!h]
%\begin{figure}
% \vspace*{-2.0 cm}
%\begin{center}
 \includegraphics[width=3.25in]{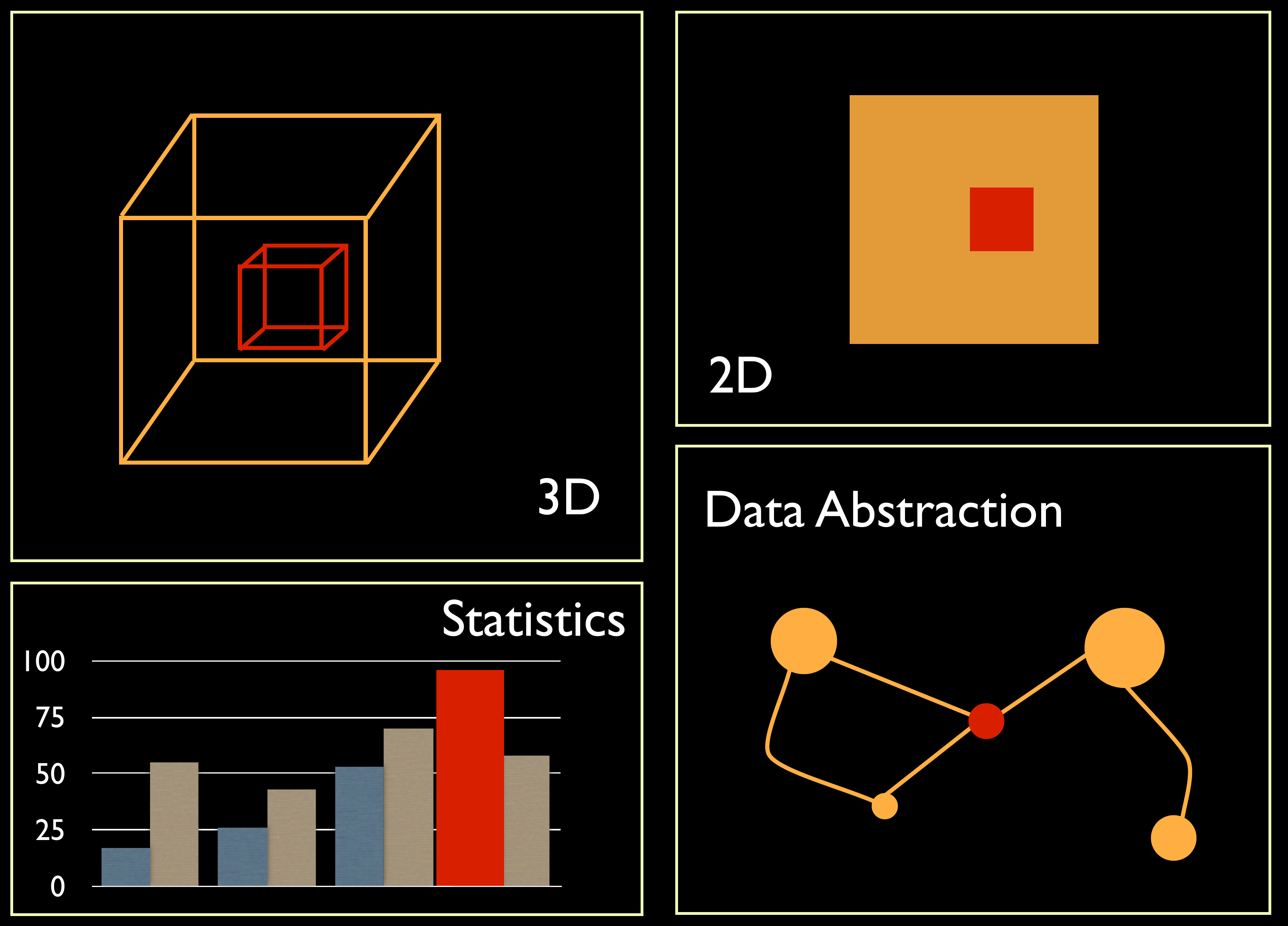} 
\vspace*{0.2 cm}
\caption{Linked Views (figure created by M. Borkin) }
\label{abstraction}
\end{figure}

Live {\it linking} of views across display modes holds the key to effective visualization and analysis of high-dimensional data sets \cite[cf.][]{Tukey1977, Wong1997, Gresh2000}.    Figure \ref{abstraction} shows a cartoon where four types of graphical display of a high-dimensional data set are pictured with one data {\it subset} highlighted in red.  In an effective ``linked view" visualization system, the kind of highlighting the red coloring represents is done {\it interactively, in real time}, and the {\it selections made can be saved} and combined with other selections for use in analysis.\footnote{see \url{www.kitware.com/InfovisWiki/index.php/Linked_Views} and references cited there for more information}

When researchers can {\it easily} investigate the behavior of trends and outliers in all the dimensions of data at hand they will learn more about the information at hand as a result.  (Conversely, if it is {\it not easy} to carry out exploratory investigations, researchers will often stop analysis at a stage  where key insights will remain hidden within a high-dimensional data set.)  For an astronomical example, imagine that a particular group of points in an $x$-$y$ plot of flux vs. velocity appeared to have aberrant behavior.  In a linked-view system, a user could immediately highlight, select, and optionally include/exclude those points from display and analysis amongst other dimensions, for example in a plot of velocity vs. signal-to-noise ratio, which might show the aberrant points to have low significance.   Extrapolating from this simple example, one can imagine and appreciate the power real-time linked views offer for making  more sophisticated investigations, such as data selection based on behavior seen in a combination of several dimensions.  In Astronomy, the ability to interactively explore the connections between data points in statistical graphs and the same measurements' positions in ``real" 3D space, and vice-versa, is particularly powerful.

In the realm of point-based data (e.g. ASCII tables), the benefits of interactive linked views were first explored by John Tukey and his colleagues using the PRIM-9 system they developed in the 1970s.\footnote{See \citet{Friedman2002} for a review.} No readily-available computers 40 years ago had input devices that could be used to graphically select subsets of data, so Tukey's team had to design a custom visualization control box with many buttons all of which had special selection- and manipulation-oriented functions.\footnote{An excellent demonstration video showing PRIM-9 is at \url{stat-graphics.org/movies/prim9.html.}}  Tukey's ideas on {\it Exploratory Data Analysis} \citep{Tukey1977}, including principles he called ``picturing," ``rotation," ``isolation," ``brushing," and ``masking," were first implemented commercially in 1986 in the Macintosh-only program {\it DataDesk}, which is still in use today on Macs and PCs.\footnote{In 1986 the Macintosh operating system, then two years old, was the only widely-available computer with a mouse-driven graphical user interface needed to make the PRIM-9 ideas practicable.}  

LIST 1 gives a summary of the mainstream commercial  descendants and offshoots of the Exploratory Data Analysis principles espoused by Tukey.  These are powerful tools for exploring tabular data on its own, but {\it none} of them links image-based or image-cube-based information to catalog (tabular) data, which is the {\bf key missing link} in astronomical data analysis today.
\vskip .1in

\noindent 
\begin{center}
{\bf LIST 1: 
Commercial Linked View Software for Analyzing Tabular Data\footnote{An interesting comparison of the last three services, and the similar ``QlikView" software (\burl{qlikview.com}) is at \burl{www.practicaldb.com/blog/data-visualization-comparison}.}
}
\end{center}
\noindent {\bf DataDesk, est. 1986}
\vskip 0.001in
\noindent \url {www.datadesk.com}, inspired by John Tukey's and Paul Velleman's work on ``Exploratory Data Analysis", see \citet {Friedman2002} for a review
\vskip 0.1in
\noindent {\bf Spotfire, est. 1996}
\vskip 0.001in
\noindent \url{spotfire.tibco.com}, inspired by Chris Ahlberg's and Ben Shneiderman's  ideas about interactive data display, see \url{www.cs.umd.edu/hcil/spotfire/} and references therein, including  \citet{Ahlberg1994}
 \vskip 0.1in
\noindent {\bf Tableau, est. 2003}  
\vskip 0.001in
\noindent \url{www.tableausoftware.com}, inspired by Chris Stolte, Diane Tang, and Pat Hanrahan's work on ``Polaris" and VizQL (Visual Query Language), see  \citet{Stolte2002} 
 \vskip 0.1in
\noindent {\bf Microsoft Business Intelligence (``BI"), est. 2000's}  
\vskip 0.001in
\noindent \url{www.microsoft.com/en-us/bi/default.aspx}, inspired by extensions to Microsoft's SQL database services and Excel spreadsheet (in the form of ``PowerPivot" add-on)
\begin{figure}[!h]
\begin{center}
 \includegraphics[width=1.4in]{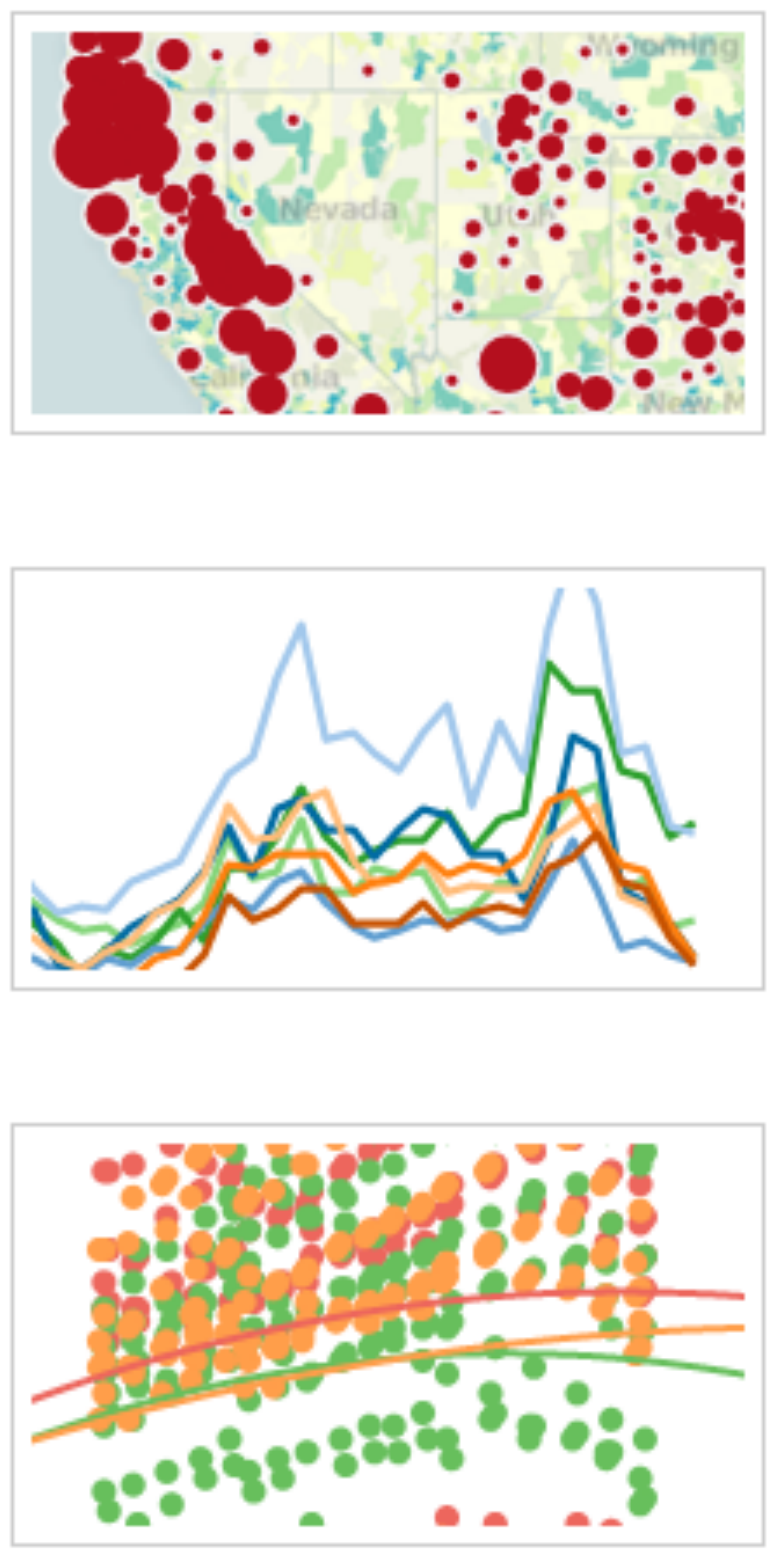} 
\vspace*{0.2 cm}
\caption{Tableau Samples}
\label{tableau}
\end{center}
\end{figure}

Figure \ref{tableau} shows a screenshot of just a few of the kinds of graphs that Tableau and its ilk produce.  Color is used to subset and link points shown in multiple displays, and subsets of particular colors can be defined graphically or algorithmically, and they can be saved.  Pre-made maps like the one shown in the top panel can be used as backgrounds and pre-defined bounded regions (e.g. US states) can be used as selection facets--but new boundaries within an image (known as new ``segmentations") {\it cannot} be easily added.

Similar linked-view software packages for exploring tabular data are available in the Open Source community (LIST 2).  These often have less intuitive or polished graphical interfaces, but they may become exceptionally useful as flexible, statistically-sophisticated, modules that can be integrated into a set of inter-operable tools as discussed in Section 4.

\vskip 0.2in
\begin{center}
{\bf LIST 2: 
Sample Open-Source and/or Free Linked View Software for Analyzing Tabular Data}
\end{center}
\vskip 0.001in
\noindent {\bf ggobi}: \url {www.ggobi.org}, cf. the "rggobi" package in R/CRAN 
\url {cran.r-project.org/web/packages/rggobi}
\vskip 0.01in
\noindent {\bf Mondrian}: \url{stats.math.uni-augsburg.de/Mondrian}
\vskip 0.01in
\noindent {\bf Weave}\footnote{Note that this ``Weave" is {\it not} the same as the WEAVE program developed at IBM and described in \citet{Gresh2000}}: \url{www.oicweave.org}
\vskip 0.01in
\noindent{\bf Viewpoints}: \url{astrophysics.arc.nasa.gov/~pgazis/viewpoints.htm}
\vskip 0.01in
\noindent{\bf XmdvTool}: \url{http://davis.wpi.edu/xmdv}
\vskip 0.01in
\noindent{\bf TOPCAT}: \url {www.star.bris.ac.uk/~mbt/topcat}
\vskip 0.01in
\noindent{\bf ViVA Workbench}: \url{http://iplant-viva.sourceforge.net/}
\vskip 0.01in
\noindent{\bf TITAN}: \url{www.kitware.com/InfovisWiki/index.php/Main_Page}

\vskip .1in 
In geography and demographics, so-called ``GIS" or ``Geographic Information System" tools such as ESRI's {\it ArcGIS}\footnote{\url{www.esri.com/software/arcgis/}} and Pitney Bowes' {\it MapInfo Professional}\footnote{\url{www.pbinsight.com/products/location-intelligence/}}  and {\it Engage3D Pro}\footnote{\url{www.encom.com.au/template2.asp?pageid=149}} offer powerful linked-view systems where maps are used as layers.  Importantly, though the maps  themselves are not typically treated as data pixel-by-pixel so that selection within a map is usually along pre-defined region boundaries, making the selection and extraction of map-based data for an arbitrary user-selected region less than fully straightforward.

So, {\it are there any working robust tools that offer image- and cube-savvy linked-view visualization and analysis environments?}   Sort of. In the early 2000's there were two notable attempts to implement an image and/or cube-enabled linked view tool: WEAVE at IBM \citep{Gresh2000} and MIRAGE at Bell Labs/Lucent \citep{2003ASPC..295..339H}.   

\begin{figure}[!h]
\begin{center}
\includegraphics[width=3.0in]{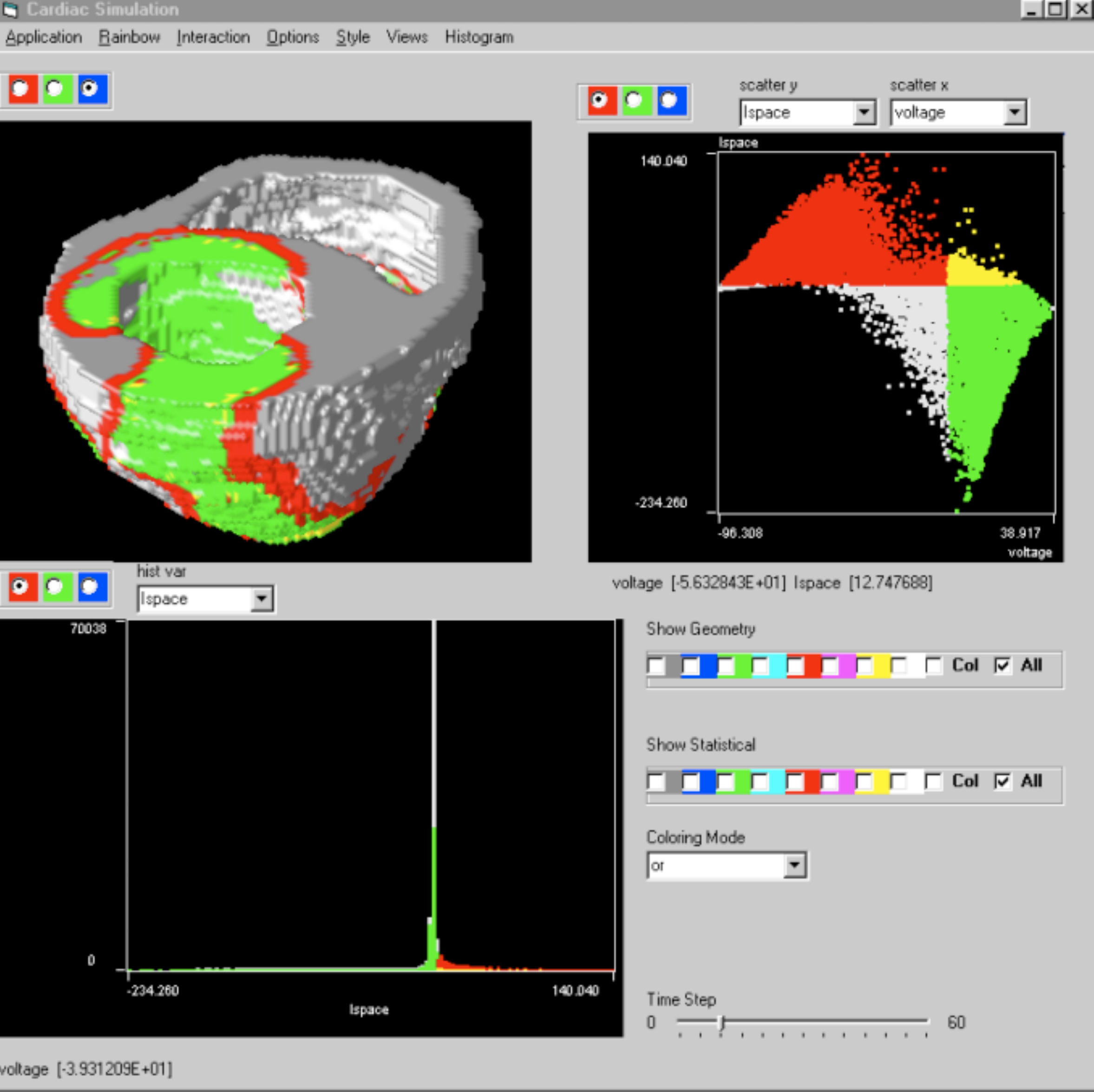} 
\end {center}
\vspace*{0.2 cm}
\caption{Screen Shot of ``WEAVE" in action.  Colored selections can be made in any of the 2D analysis panels, or directly in the high-dimensional (3D) display, and all views are live-linked. From \citet{Gresh2000}. }
\label{weave}
\end{figure}

WEAVE was developed in Bernice Rogowitz' group at IBM Research, to support a collaboration between computer and cognitive scientists with medical researchers.  It comes the closest to a system that would be perfect for the analysis of high-dimensional astronomical data (see Figure \ref{weave}).  Unfortunately, though, the IBM WEAVE project's software, built linking Data Explorer and Diamond (a precursor to Opal/ViVA, see LIST 2) via ActiveX, is no longer supported or available.

MIRAGE was developed by visualization and statistics researchers collaborating with astronomers, directly for use in Astronomy, and it has been integrated to some extent with Virtual Observatory standards \citep{2004ASPC..314..300C, HoT.K.2007}.  As of this writing, MIRAGE can be acquired at \url {skyservice.pha.jhu.edu/develop/vo/mirage/}, but its VO functionality is presently somewhat fragile.  Furthermore, MIRAGE does not allow region selection (segmentation) within images like WEAVE did, and it does not presently handle data cubes.

In spite of their limitations, WEAVE and MIRAGE demonstrated the potential of exploratory data analysis tools that understand 2D and 3D images.   Yet, since WEAVE is no longer available, and MIRAGE's image-based-information linking is limited, neither offers a full linked-view solution to astronomical researchers today.  More recent efforts built on top of visualization toolkits like VTK (discussed below), are presently extending the image-enabled linked-view paradigm that WEAVE and MIRAGE pioneered.

At present, choices open to astronomers seeking to implement high-dimensional linked-view visualization and analysis into their research can be categorized into the four kinds of approaches itemized in LIST 3.  Examples given in the list are discussed in turn, below.
\begin{figure*}[!h]
\begin{center}
\includegraphics[width=5.7in]{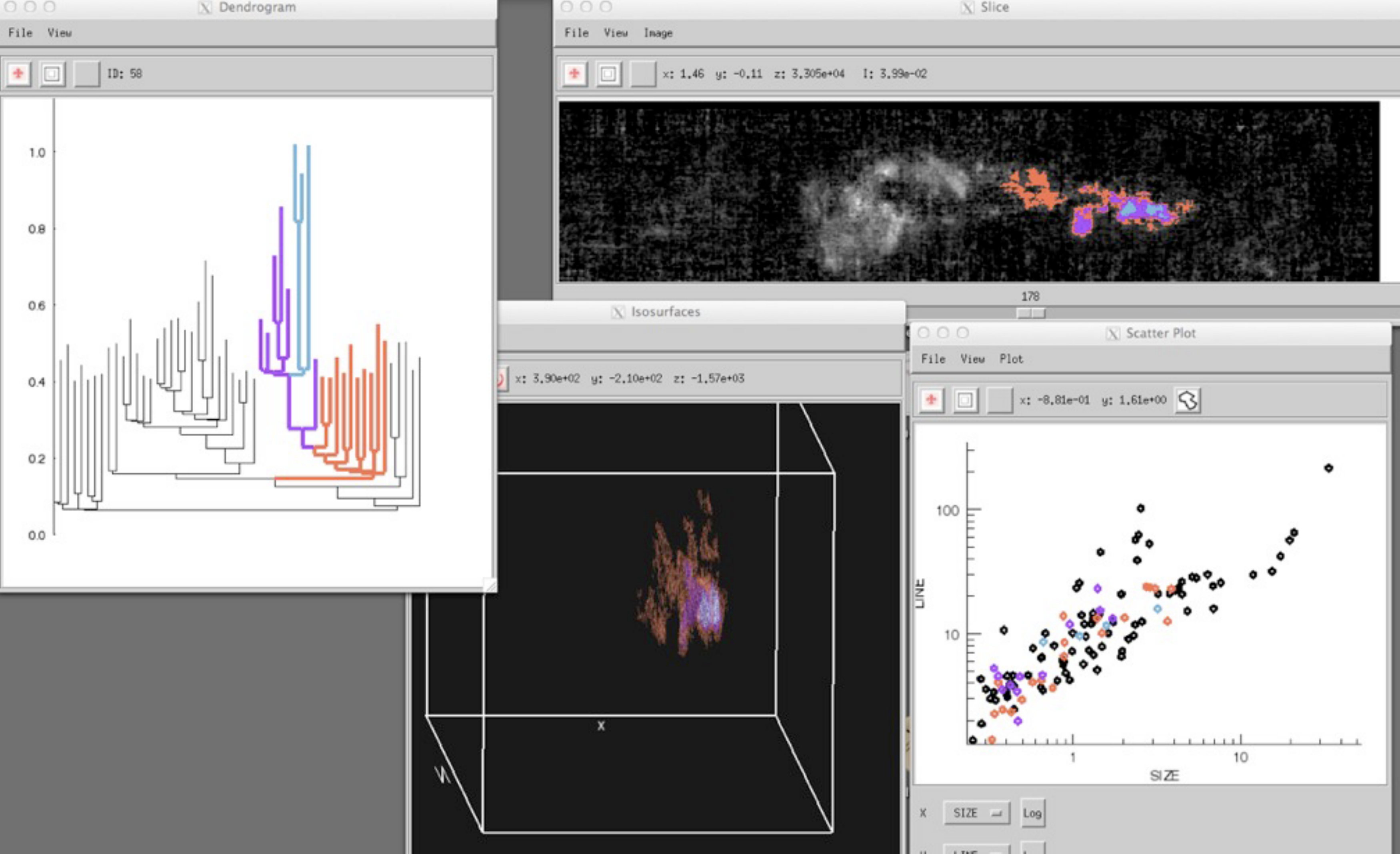} 
\end {center}
\vspace*{0.2 cm}
\caption{Example: Screen Shot of ``DendroViz" project, courtesy C. Beaumont. }
\label{cloudviz}
\end{figure*}

\begin{center}
\noindent
{\bf LIST 3: Approaches to High-Dimensional Image- and Cube-Aware Linked View Visualization in Astronomy}
\end{center}
\begin{enumerate}
\item Use existing high-level  visualization and analysis packages that satisfy astronomy-specific requirements, such as IDL, to implement custom linked-view tools for specific problems.  {\it Example:} Dendroviz.
\item Use resource-hub and/or message-passing architectures to inter-connect software packages in a way that they can link their views to a limited extent. {\it Example:}  SAMP.
\item Adapt capabilities from software systems from beyond Astronomy. {\it Example:} Astronomical Medicine.
\item Build a new extensible system, preferably based on open-source, re-usable, modules.  {\it Examples:}  Glue, Paraview, Titan.
\end{enumerate}

\subsection{Custom Solutions within Existing Software, {\it e.g. Dendroviz}}

The screenshot in Figure \ref{cloudviz} shows a linked-view display of a spectral-line data cube.  The ``Dendroviz" (a.k.a. ``Cloudviz") software used to create the views was written inside of IDL.\footnote{\url{www.exelisvis.com/ProductsServices/IDL.aspx}} It is freely available\footnote{at \url{code.google.com/p/cloud-viz/}} to IDL users, and was written by Ph.D. student Christopher Beaumont for his thesis work at Harvard.   Many of the desirable aspects of linked views discussed above, and schematized in Figure \ref{abstraction}, are incorporated here.  The tree-like diagram at upper left in the figure shows a hierarchical decomposition of the spectral-line intensity within a 3D (position-position-velocity) cube.  The $x$-$y$ plot at lower right shows another physical diagnostic of the gas, and the two other panels show volume visualizations and slice views of the data.  Linking is possible by selecting in any 2D analysis plot (e.g., tree, $x$-$y$) and then seeing selections as colored regions within the 2D (slice) and 3D (volume) data displays.  Selections can be saved, combined, and output as filters.\footnote{Videos demonstrating Dendroviz functionality and usage are online at \url{projects.iq.harvard.edu/seamlessastronomy/software/dendrograms.}} 

Thus, it is possible within a general-purpose program like IDL to design a custom linked-view environment.  But, this approach has some serious limitations.  First, it is difficult or impossible to make arbitrarily-shaped selections within the image-based environment.  And second, the functional and aesthetic qualities of the user interface and visualization layouts here are not very good, and they cannot be improved when one is restricted to using only IDL.

\subsection{Hubs and Message Passing Amongst Disparate Programs, {\it e.g. SAMP}}

A much more general approach to linking views of astronomical data is offered by SAMP, a message-passing architecture developed by Mark Taylor and colleagues within the International Virtual Observatory Community.\footnote{The SAMP standard is described at \url{www.ivoa.net/Documents/SAMP}}  

\begin{figure*}[!h]
\begin{center}
\includegraphics[width=5.8in]{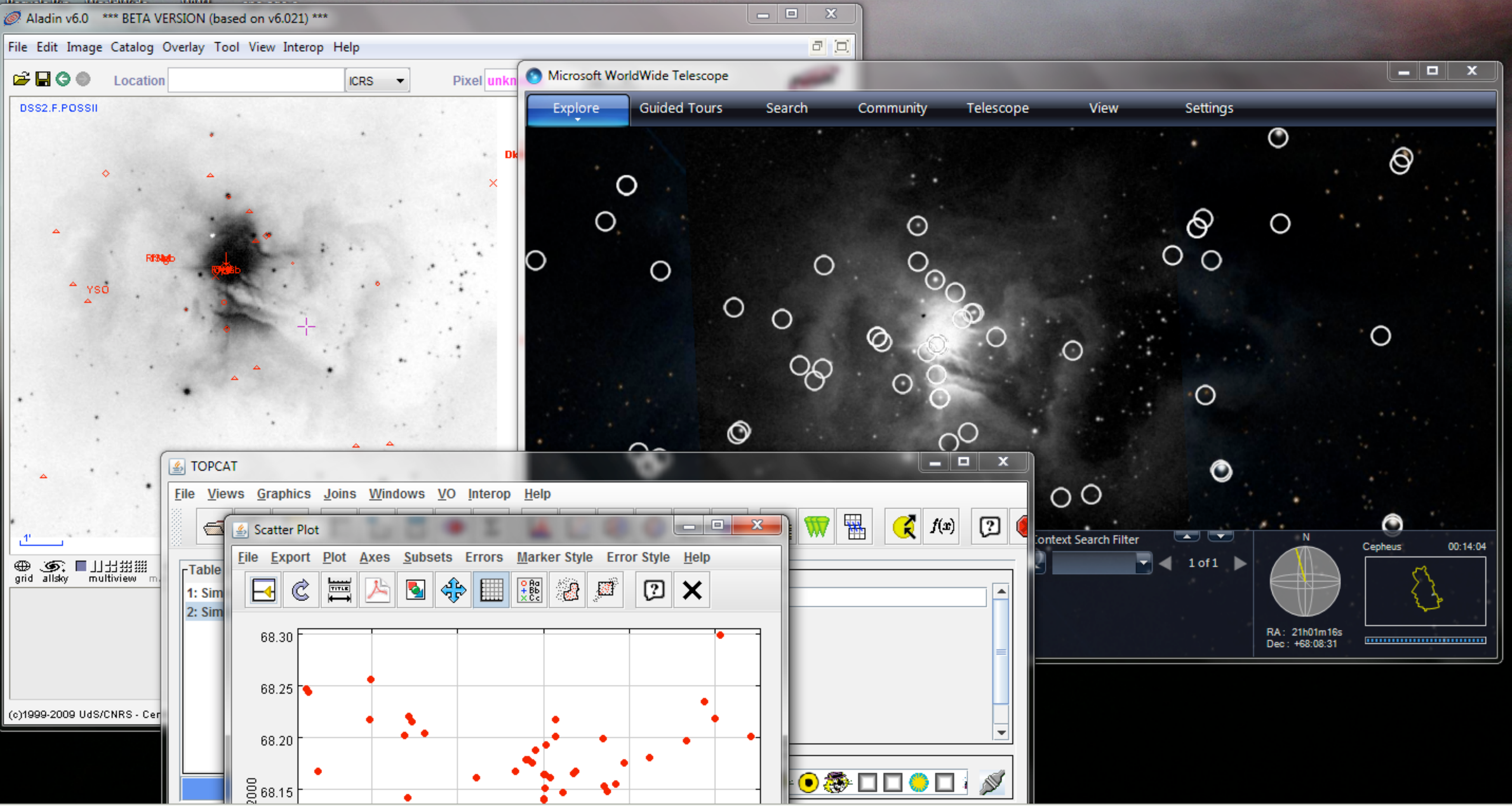} 
\end {center}
\vspace*{0.2 cm}
\caption{Example: Screen Shot of SAMP-connected Applications }
\label{samp}
\end{figure*}

Figure \ref{samp} shows a screen shot of SAMP in action.  At the upper left, an Aladin\footnote{\url {aladin.u-strasbg.fr}} window is open showing the cluster NGC7023, with several catalog sources overlaid.  The same region of the sky and catalog data are shown in WorldWide Telescope\footnote{\url {worldwidetelescope.org}} (upper right), and the catalog data are shown in TOPCAT\footnote{\url {www.star.bris.ac.uk/~mbt/topcat}}, which can manipulate those data in a statistical-graphics environment not unlike DataDesk (lower left).  Other popular astronomical analysis environments, like ds9\footnote{\url {hea-www.harvard.edu/RD/ds9},} can also connect to SAMP, but are not shown in this example. 

So, what does SAMP do?  When applications ``connect" to the SAMP hub, as they were during the session captured in Figure \ref{samp}, they pass simple messages amongst themselves, telling each other what coordinates and field of view are currently being used, and what catalog sources are selected and sub-setted.  Thus, a savvy user can run SAMP to effectively link views, bringing the functionality of several programs to bear on the same data set at once, in a concerted way.  Other than the screen real-estate challenge posed by the need to keep track of the (many!) windows open while SAMP connects disparate applications, the major limitation of the SAMP system at present is the lack of tools to select  arbitrary regions within an image, and link such selections. 

The good news is that SAMP has recently been web-enabled, so that java and web-based applications can now be connected within a fully online environment.  

\subsection{Adaptation from Beyond Astronomy, {\it e.g. Astronomical Medicine}}

Astronomy is not the only field faced with the challenge of incorporating high-dimensional information into quantitative analyses: geography, medicine, biology, and other fields share similar challenges. The overlap of methods used in these fields,  especially in astronomical and medical imaging and analysis, is far greater than one might imagine at first.   Over the past five years, a group of us at Harvard\footnote{\url {am.iic.harvard.edu}} have been exploring the efficacy of directly adapting tools developed for medical imaging into the astronomical research environment \citep[e.g.][]{2007ASPC..376..621B}.  

\begin{figure}[!h]
\begin{center}
\includegraphics[width=2.8in]{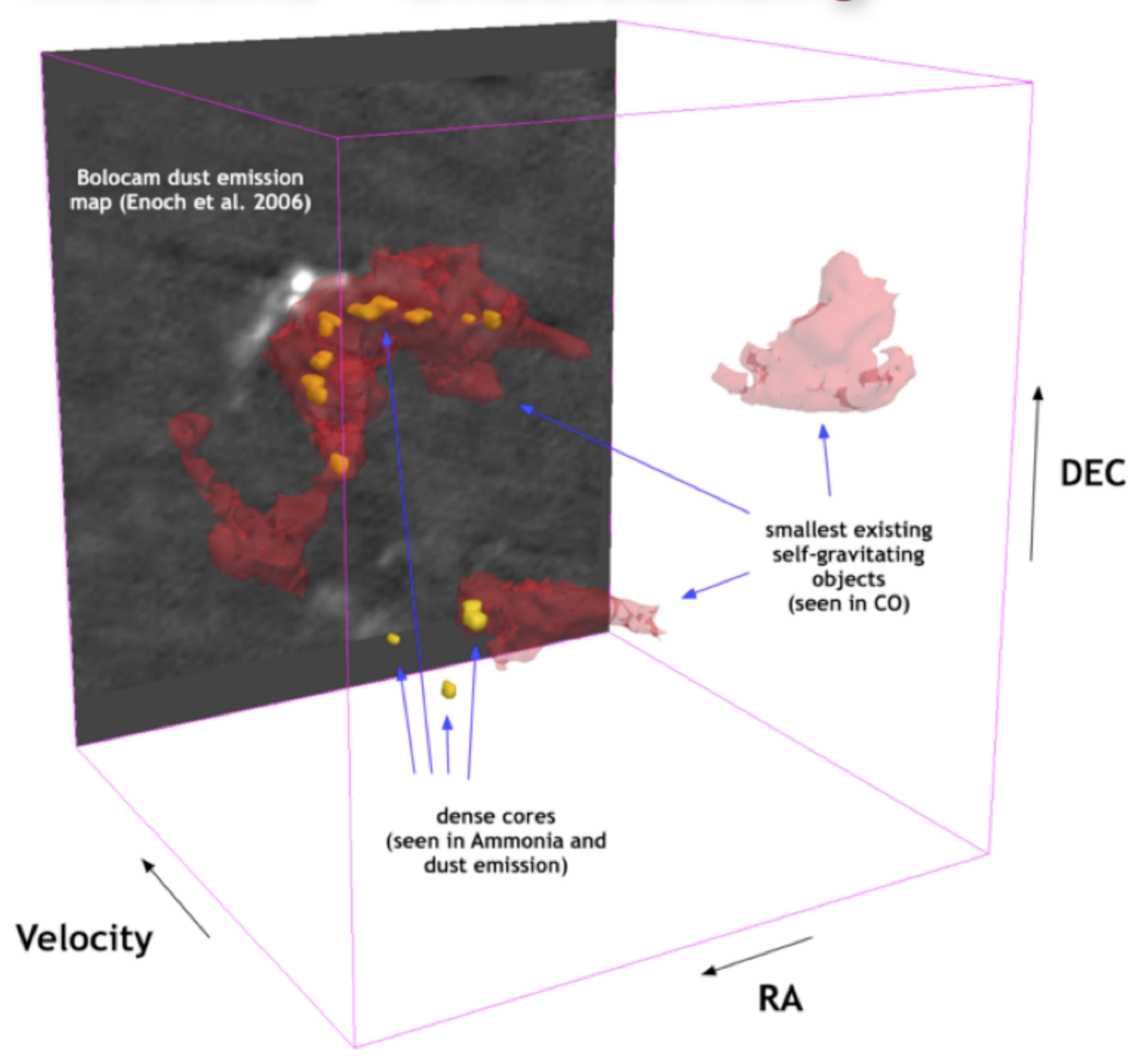} 
\end {center}
\vspace*{0.2 cm}
\caption{``Astronomical Medicine" view of L1448, created using 3D Slicer by Jens Kauffmann.  Similar figures were published as the first interactive 3D PDFs in the journal {\it Nauture} \citep{Goodman2009a}. }
\label{astromed}
\end{figure}

It is clear the the high-dimensional visualization and manipulation tools available in the medical community, largely based on the VTK and ITK toolkits (discussed further in \S3.4), are far superior to those typically available to astronomers.  Figure \ref{astromed} shows an example of the use of 3DSlicer, a program developed in part at the Surgical Planning laboratory of Brigham and Women's Hospital in Boston, used to view data about a star-forming region.  Notice that Figure \ref{astromed}  shows multiple 3D spectral-line data sets at once, and a moveable (black and white) 2D plane showing a 2D dust image is incorporated as well.   Our group at Harvard's Initiative in Innovative Computing  managed to write a converter (fits2itk) aware of astronomical coordinate systems to move FITS images into the ITK format\footnote{available at \url{am.iic.harvard.edu/FITS-reader}}, but preserving more than astronomical metadata beyond coordinates was not trivial in the medically-optimized 3D Slicer environment.

Most importantly, while medical tools can offer great visualizations, they do not typically implement linked views of tabular data that can interact with  volume- and slice-based visualizations.  Thus, for now, it is necessary to separate high-performance visualization work from statistical analyses using medically-optimized systems, but we expect that situation to change in the near future--and we look forward to trying out more software developed for other fields within the astronomical context.

\subsection{New Solutions in Open Source Environments, {\it e.g. Glue, Paraview, Titan}}

More than a decade ago \citet{North2000} investigated the idea that non-programing users could ``snap together" visualization modules on-the-fly to create whatever custom linked-view environment would best address a particular problem.  The Dendroviz solution discussed above is an implementation of this approach within IDL, but it requires a programming-savvy user.    

With the ascension of python as the preferred modern programming language within Astronomy and other fields of science, there has been an explosion in the amount of open-source code available to researchers for re-use.  (See \url {www.astropython.org}  and  \url{www.scipy.org}.)  Several graphics and table-manipulation packages are already available, and many of them even understand astronomical coordinate systems and units(!).  Similarly, the statistical analyses available within the R language (and the accompanying CRAN packages\footnote{\url{http://cran.r-project.org}}) can be interconnected to create nearly any needed analysis.   If these packages can be ``glued" together effectively, then it should be possible, even for non-programming users, to create a linked-view visualization and analysis environment using primarily free python-based and R-based\footnote{The Rpy tools at \url{rpy.sourceforge.net} allow R functions to be accessed from within python.} tools in the very near-term future.   

A group of us (Christopher Beaumont, Michelle Borkin, Thomas Robitaille, Hanspeter Pfister and me) are actively working on a new python-based linked-view visualization system code-named ``Glue."  We have already created a hub that allows various python modules to be ``linked" without their code being merged.  Currently, we are working on the user interface, and ultimately our plan is to connect Glue to R and to SAMP-enabled applications, which would offer a true ``snap together" linked-view visualization environment for Astronomers, and for other researchers.   Glue will be fully open-source, available , and we more than welcome collaboration from the community in this endeavor.\footnote{See \url {projects.iq.harvard.edu/seamlessastronomy/software} for more information.}

Glue should be able to build upon and extend important packages that use the Visualization Tool Kit (VTK)\footnote{\url{www.kitware.com/products/books.html}} as a scientific visualization platform.  The 3DSlicer program used in the Astronomical Medicine project is an example VTK being used to create a sophisticated medical visualization system. More general efforts, such as Paraview,\footnote{\url{www.paraview.org}} are applicable to many non-medical data formats, including astronomical simulation outputs (but not yet observational formats that use astronomical coordinates).  And, most promisingly, the collaborative Titan effort,\footnote{\url{www.kitware.com/InfovisWiki/index.php/Main_Page}} marries VTK-based scientific visualization to information and statistical visualization modules,\footnote{There is a growing class of such efforts, including the Mayavi project \citep{Ramachandran2011}, which combines VTK and Python in a modular, extensible, fashion.} including those from open-source efforts such as R/CRAN.

\section{Seamless Astronomy: A Vision for the Future}

To understand what we\footnote{\url{projects.iq.harvard.edu/seamlessastronomy/}} mean when we say we strive for ``seamless" astronomical research, imagine this:\footnote{Footnotes in this section offer live links to what is  possible now.}

\begin{figure*}[!h]
\begin{center}
 \includegraphics[width=5.7in]{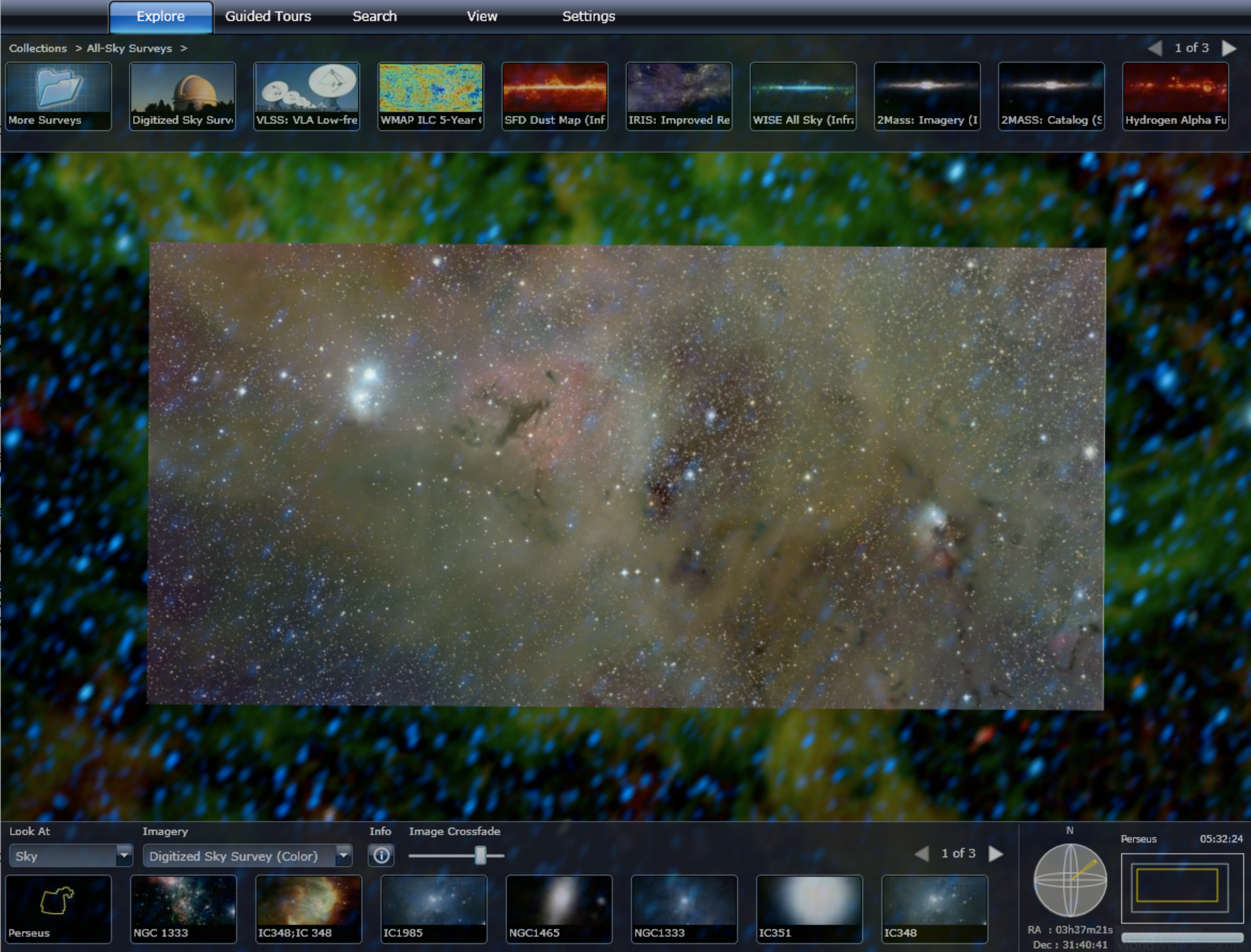} 
\vspace*{0.2 cm}
\caption{Perseus image displayed in WWT via astrometry.net and flickr.  (Foreground=wide-field optical image that was uploaded to flickr's astrometry group; Background=WISE all-sky survey). }
\label{wwt}
\end{center}
\end{figure*}

\noindent
A smartphone application featuring interesting new astronomy images shows you the inset image in the middle of Figure \ref{wwt}.   You have wireless connectivity and some kind of large display handy, and you are curious to know more.  First, you flick the image off your phone to your large display.\footnote{Already possible using, for example, AirPlay  from Apple.}  Next, you find out where this image belongs on the sky, using a recognition service that either examines embedded metadata in its header\footnote{Possible using AVM tags, see \url {virtualastronomy.org/avm_metadata.php}.}, or its content\footnote{\url{astrometry.net} can find the position of any image just based on the pattern of visible stars it contains}.  Now you use VO services embedded in any number of applications, for example the WorldWide Telescope\footnote{\url{worldwidetelescope.org}}\citep{Goodman2012c}, to put this image in context, allowing you to view how it looks in comparison to extant images at many wavelengths.\footnote{Presently in WWT, one can locate an image based on a FITS header, AVM header, from metadata passed from astrometry.net \citep{Lang2010}, via flickr \url{www.flickr.com/groups/astrometry/}, or by registering features by hand. The WWT view shown in Figure \ref{wwt} can be recreated at \url{tinyurl.com/seeperseus}, by zooming out a bit and then selecting WISE from the Collection ``All Sky Surveys" as the background imagery.} Figure \ref{wwt} shows the result of uploading this image to the ``astrometry" group on flickr, and then selecting the ``View in WorldWide Telescope" link that appears in the comments on the resulting page\footnote{\url{www.flickr.com/photos/66496709@N00/6791649829/}} a few minutes later, and then changing the background view to show the latest WISE infrared imagery.  You're wondering about the young-star population in the area, so you first use the VO-searching capabilities built in to WWT to add an overlay of 2MASS sources (not shown here, due to high density of such sources!), and then later you connect WWT to other astronomical and visualization applications using SAMP and Glue. You're curious if there are any molecular-line maps of this region, so you use the features built-in to WWT and ADSLabs\footnote{ \url{adslabs.org}} to find and display a list of all the papers that  mention ``data cubes" and study this region.  One of them has a great map of $^{13}$CO emission in the Perseus, and you want to see the 2D images in Figure \ref{wwt} and the catalogs you have retrieved online in the context of those 3D maps.  You're lucky and the person publishing the CO map included persistent hdl tags in her paper that lead you to a ``Dataverse"\footnote{\url{thedata.org}} online repository at \url {theastrodata.org},  where you can retrieve and/or link to the data cube.\footnote{at \url{tinyurl.com/complete13COper}} Now, you call upon the capabilities of Glue to display and analyze a live-linked combination of: the 3D spectral-line data; moveable planes that hold the imagery shown in Figure \ref{wwt};  the catalogs you've linked to via SAMP; and a calculated ``dendrogram" decomposition of the 3D data you calculated using a module within Glue.  

Using exploratory data analysis and linked views, you begin to notice correlations and outliers amongst the various dimensions of data you've displayed.  It's tricky to make and explore some of the selections you want to make within the 3D volumes, so you use your hands in the air, as sensed by a high-dimensional pointing device,\footnote{At present, the Microsoft Kinect is a good, albeit low-resolution, example of such a device. The Leap \url {www.leapmotion.com} may be the next, higher-resolution step.} to make those selections. It seems that there are big shells within the CO data set that seem associated with young stars.  How young are the stars?  You go to the \url {astrobetter.com} site and discover a new algorithm, written in R, that offers better estimates of young stars' ages.  So, you download that algorithm, and you kindly decide to make this new algorithm part of Glue by using  Rpy, the Python interface to R,\footnote{\url{rpy.sourceforge.net/}} to create a small Python program that uses R's statistical power to analyze the information about the young stars.  When you're done with your analysis, you: 
\begin{enumerate}
\item publish your new Python-based young star age module to the Glue code repositories online (e.g. Github,  Sourceforge)
\item publish a paper in a Journal about your findings, including persistent identifiers to the data used in each graph and/or analysis shown (e.g. using the Dataverse architecture at theastrodata.org)
\item include interactive figures in your Journal article, and on your web site, allowing others to explore your data further (similar to the interactive 3D PDF published by \citet{Goodman2009a} in {\it Nature} in 2009), which you create following the free instructions by Josh Peek posted on astrobetter.com.\footnote{available at \url{tinyurl.com/peek3dpdf}}

As the footnotes demonstrate, about 90\% of this scenario is  possible now, even though astronomers are not typically aware of all of the tools that make it possible.  It's the last 10\%, which includes implementing a Glue-like solution and creating effective 3D interaction techniques, that stands between now and a seamless future of fully linked-views in high-dimensional visualization.
\end{enumerate}

\section{Challenges}

Visualization researchers have been working to optimize high-dimensional linked-view visualization for nearly 40 years.  Many systems exist that are great for point-based data, but this paper demonstrates that none of these systems yet addresses astronomical image- and cube-based data sets adequately.    The main challenges to implementing a system for viewing, manipulating, and inter-comparing high-dimensional astronomical data sets in a linked-view environments at present are:

\begin{enumerate}
\item {\bf Big data}--today's laptops can easily handle data sets like any of the ones used as examples in this paper.  But, instruments like ALMA and integral field units, and big numerical simulations, generate data sets far too large to manipulate within current computing architectures.  Additional research is needed on how to most effectively retrieve and load subsets of information into a computer's memory, so as to still allow real-time exploration and manipulation of even the largest data sets.  Clever structuring of data sets using new databases like SciDB\footnote{\url {www.scidb.org}}, and the continued evolution of MapReduce and Hadoop may help, but work will still be needed to optimize remote ``real-time" access to subsets of  very large data sets.
\item {\bf Interface design}--it is difficult to avoid complicated menus and too many open windows.  The cognitive load a very flexible system places on a user can be much greater than a more rigid system, so ``snap together" customizable tools can ultimately cause confusion if implemented poorly.\footnote{see \citet{BerniceE.RogowitzaandNaimMatasci}}  Thus, it will be critical to study a range of user interface options, and match those options well to user's needs, and to their equipment.  Ten open windows may be fine if one has a giant monitor, touch table, and/or display wall, but a system that requires all those windows will not likely work well on portable devices. It is also critical, and difficult, to design a system that best supports problem solving without overwhelming a user with options.
\item {\bf 3D selection}--mice, trackpads, and touch screens have evolved over the past 30 years to offer very good options for selecting regions of a two-dimensional screen.  But, research into 3D selection has barely begun.\footnote{See the work of Daniel Keefe et al. for good examples of what's presently possible, at  \url {ivlab.cs.umn.edu/project_3dui.php}.}  Human hands cannot be moved as steadily in 3D free space as they can be on 2D surfaces, so while Kinect-like devices offer inroads, they may not immediately offer optimal solutions.
\item {\bf Diversity of challenges}--the examples of astronomical research challenges used in this paper are but a tiny fraction of the range of problems researchers will bring to a system for linked-view visualization of high-dimensional data.
\end{enumerate}

I predict that these four challenges, and others not yet anticipated, will be met through a combination of three trends that we can see emerging already.  \begin{enumerate}
\item {\bf Modularity}--As mashup-style software solutions become more and more prevalent on the web today, there is every reason to expect that an approach where expert-developed modules, each aimed at addressing a particular needs, can be ``glued" together effectively if appropriate attention is paid to standards and compatibility.
\item {\bf Open source collaborative software}--The number of astronomers, visualization researchers, and generally generous coders who seem interested in helping to develop useful code for research and visualization is constantly growing.  This growth combined with the increasing ease with which coders can share their work, thanks to platforms like Github, Google Code, and Sourceforge, should make it possible for an ever-widening range of talent and ideas to be brought to bare on these challenges.
\item {\bf Interdisiplinary collaboration}--The Astronomical Medicine project offers just one demonstration that the need for linked-view visualization of images and data cubes is shared across fields.   As the number of fields faced with high-dimensional visualization challenges expands, so will funding for and work on this problem.
\end{enumerate}
Thus, it appears that the time is ripe for Astronomers to collaborate  beyond our field's traditional boundaries in order to create a modular open-source high-dimensional linked-view exploratory data visualization environment.

\acknowledgements
The author thanks her collaborators Michelle Borkin and Christopher Beaumont for their significant contributions to this work, and Bernice Rogowitz and Hanspeter Pfister for excellent suggestions on improving it.  Microsoft Research, the National Science Foundation, and NASA all fund the author's work on astronomical data visualization.

\newpage%%%%%%%%%%%%%%%%%%%%%%%%%%%%%%%%%%%%%%%%%%%%%%%%%%%%%%
%\bibliography{/Users/agoodman/Documents/AstroVizTools.bib}{} You commented this out for astro-ph compliance.
\bibliography{AstroVizTools.bib}{}
\bibliographystyle{apj}

\end{document}